\documentclass[aps,prl,reprint,superscriptaddress]{revtex4-1}
\usepackage{graphicx}
\usepackage{amssymb,amsmath}
\usepackage{psfrag}
\usepackage{color}

\newcommand{\iu}{\mathrm{i}}	
\newcommand{\abs}[1]{\ensuremath{\left| #1 \right|}}
\newcommand{\evalat}[2]{\left. {#1} \right|_{#2}}
\DeclareMathOperator{\im}{Im}
\DeclareMathOperator{\re}{Re}

\newcommand{\pdiffn}[3]{\frac{\partial^{#3}#1}{\partial #2^{#3}}}

\newcommand{\bra}[1]{\ensuremath{\left\langle #1 \right|}}
\newcommand{\ket}[1]{\ensuremath{\left| #1\right\rangle}}
\newcommand{\braket}[2]{\ensuremath{\left\langle #1 | #2\right\rangle}}
\newcommand{\melement}[3]{\ensuremath{\left\langle #1\left| #2 \right| #3\right\rangle}}

\begin{document}
\title{Attosecond streaking enables the measurement of quantum phase}

\author{V. S. Yakovlev}
\email{vladislav.yakovlev@physik.uni-muenchen.de}
\affiliation{Department f\"ur Physik, Ludwig-Maximilians-Universit\"at, Am Coulombwall 1}
\affiliation{Max-Planck-Institut f\"ur Quantenoptik, Hans-Kopfermann-Stra{\ss}e 1, D-85748 Garching, Germany}

\author{J. Gagnon}
\affiliation{Department f\"ur Physik, Ludwig-Maximilians-Universit\"at, Am Coulombwall 1}
\affiliation{Max-Planck-Institut f\"ur Quantenoptik, Hans-Kopfermann-Stra{\ss}e 1, D-85748 Garching, Germany}

\author{N. Karpowicz}
\affiliation{Max-Planck-Institut f\"ur Quantenoptik, Hans-Kopfermann-Stra{\ss}e 1, D-85748 Garching, Germany}

\author{F. Krausz}
\affiliation{Department f\"ur Physik, Ludwig-Maximilians-Universit\"at, Am Coulombwall 1}
\affiliation{Max-Planck-Institut f\"ur Quantenoptik, Hans-Kopfermann-Stra{\ss}e 1, D-85748 Garching, Germany}

\begin{abstract}
  Attosecond streaking, as a measurement technique, was originally conceived
  as a means to characterize attosecond light pulses, which is a
  good approximation if the relevant transition matrix elements are
  approximately constant within the bandwidth of the light pulse. Our analysis
  of attosecond streaking measurements on systems with a complex response to
  the photoionizing pulse reveals the relation between the momentum-space
  wave function of the outgoing electron and the result of conventional
  retrieval algorithms. This finding enables the measurement of the quantum
  phase associated with bound-continuum transitions.
\end{abstract}
\maketitle

The absorption of an energetic photon by an atom or molecule starts a sequence
of events which may result in the emission of one or several electrons. Recent
advances in attosecond science allow time-resolved measurements of such
dynamics \cite{Drescher_Nature_2002,Swoboda_PRL_2010,Schultze_Science_2010}.
In general, the resulting electron wave packets carry
valuable information about the processes that produced them. Until recently, the
full characterization of an electron wave packet was impossible---while the
spectrum of a wave packet can easily be measured, the phase information was
inaccessible. In this Letter, we show rigorously that attosecond streaking
\cite{Itatani_PRL_2002,Kitzler_PRL_2002} is an appropriate tool to measure the
energy dependence of the phase associated with a particular bound-free
transition.

Attosecond streaking consists in recording a set of photoelectron spectra over
a range of delays between an ionizing extreme ultraviolet (XUV) pulse and an
optical waveform, the intensity of which should be too weak to affect bound
electrons, but strong enough to significantly accelerate or decelerate free
electrons. This interaction with free electrons is referred to as
\emph{streaking}, and the role of the streaking waveform is usually played by
a few-cycle near-infrared laser pulse with a stabilized carrier-envelope
phase. A set of such laser-dressed electron spectra comprises a
\emph{spectrogram}.
For a comprehensive review of attosecond streaking measurements
and reconstruction techniques, we refer the reader to the original papers
\cite{Itatani_PRL_2002,Kitzler_PRL_2002,Kienberger_Nature_2004,Quere_JMO_2005,
  Gagnon_APB_2008,Gagnon_OE_2009} and focus on the most
important concepts. In the following, the photoionizing radiation is referred
to as an ``XUV pulse'', although this radiation may consist of several pulses,
and it does not have to be strictly in the XUV spectral range. Let us consider
the interaction of a linearly-polarized XUV pulse with a quantum-mechanical
system, which is initially found in a stationary bound state $\ket{\Psi_0}$ of
its Hamiltonian $\hat{H}_0$. The XUV pulse launches a single-electron wave
packet $\ket{\Psi(t)}$, the propagation of which in the ionic potential is
conveniently described in a basis formed by continuum eigenstates of
$\hat{H}_0$. In the following, $\ket{\mathbf{p}}$ will represent such an
eigenstate, so that $\abs{\braket{\mathbf{p}}{\Psi(t)}}^2$ is the
probability density of detecting a photoelectron with an asymptotic momentum
$\mathbf{p}$.
In this basis, the motion of the photoelectron is determined by
  $\braket{\mathbf{p}}{\Psi(t)} = \tilde{\chi}(\mathbf{p}) e^{-\iu \frac{p^2}{2} t}$
(atomic units are used throughout this paper). The probability amplitudes
$\tilde{\chi}(\mathbf{p})$ fully describe the properties of the electron wave
packet. For each direction of observation, we define a \emph{time-domain wave
  packet} as
\begin{multline}
  \label{eq:time_domain_WP}
  \chi(t) = \frac{\iu}{\pi}
  \int_0^\infty \tilde{\chi}(\sqrt{2\epsilon})
  e^{-\iu (\epsilon-\epsilon_0) t} \,d\epsilon =\\=
  \frac{\iu}{\pi}
  \int_0^\infty \braket{\mathbf{p}}{\Psi(t)}
  e^{\iu \frac{p_0^2}{2} t} p \,dp,
\end{multline}
where $\epsilon=p^2/2$ stands for the energy of the electron infinitely far
from the ion, and $\epsilon_0=p_0^2/2$ is a central energy of the wave
packet. As we show below, $\chi(t)$ is the quantity that is recovered by
analyzing the attosecond streaking spectrogram. This provides much needed
rigor to what has been vaguely referred to as the reconstructed
``wave packet''. 

If an electron is freed as a result of single-photon ionization, the
properties of the electron wave packet derive from the spectral components of
the incoming light pulse multiplied by the respective complex-valued
transition matrix element $D(\mathbf{p})$ \cite{Mauritsson_PRA_2005}.
Let us consider an XUV pulse with the electric field $E_\mathrm{XUV}(t) =
\re\left[\mathcal{E}_\mathrm{XUV}(t) e^{-\iu \Omega t}\right]$, where $\Omega$
is the central frequency, and $\mathcal{E}_\mathrm{XUV}(t)$ is the complex
envelope of the pulse.  First-order perturbation theory combined with the
dipole and rotating-wave approximations yields
\begin{equation}
  \label{eq:D}
  \tilde{\chi}(\mathbf{p}) = -\frac{\iu}{2}
  \tilde{\mathcal{E}}_\mathrm{XUV}\left(\frac{p^2}{2}-\frac{p_0^2}{2}\right)
  D(\mathbf{p}),
\end{equation}
where
\begin{equation}
  \label{eq:XUV_spectrum}
  \tilde{\mathcal{E}}_\mathrm{XUV}(\omega) =
  \int_{-\infty}^\infty \mathcal{E}_\mathrm{XUV}(t) e^{\iu \omega t}\,dt
\end{equation}
is the Fourier transform of the complex XUV envelope. The central momentum
$p_0$ is related to the central frequency $\Omega$ of the XUV pulse and the
ionization potential $W$ by energy conservation: $p_0^2/2 = \epsilon_0
= \Omega - W$.

In the absence of the streaking field, an explicit expression for the
photoelectron spectrum
$S_0(\mathbf{p})=\abs{\braket{\mathbf{p}}{\Psi(t)}}^2=
\abs{\tilde{\chi}(\mathbf{p})}^2$
is
\begin{equation}
  \label{eq:unstreaked_spectrum}
  S_0(\mathbf{p}) = \left| \frac{1}{2}
    \int_{-\infty}^\infty dt\, \mathcal{E}_\mathrm{XUV}(t)
    D(\mathbf{p})
    e^{\iu\left(\frac{p^2}{2}-\frac{p_0^2}{2}\right) t} \right|^2.
\end{equation}

As it was originally shown in \cite{Kitzler_PRL_2002}, the presence of a
streaking laser pulse $\mathbf{E}_\mathrm{L}(t)=-\partial
\mathbf{A}_\mathrm{L} / \partial t$ delayed by $\tau$ with respect to the XUV
pulse is accounted for by the following generalization of
Eq.~\eqref{eq:unstreaked_spectrum}:
\begin{equation}
  \label{eq:spectrogram1}
  S(\mathbf{p},\tau) = \left| \frac{1}{2}
    \int_{-\infty}^\infty dt\, \mathcal{E}_\mathrm{XUV}(t+\tau)
    G_0(\mathbf{p},t) e^{\iu\left(\frac{p^2}{2}-\frac{p_0^2}{2}\right) t} \right|^2.
\end{equation}
Here,
$  G_0(\mathbf{p},t) = 
  D\bigl(\mathbf{p}+\mathbf{A}_\mathrm{L}(t)\bigr)
  e^{\iu \Phi(\mathbf{p},t)}$
with $\Phi(\mathbf{p},t)$ being the Volkov phase:
\begin{equation}
  \label{eq:Volkov_phase}
  \Phi(\mathbf{p},t) = -\int_t^\infty
  \left(\mathbf{p} \mathbf{A}_\mathrm{L}(t') +
  \frac{1}{2} A_\mathrm{L}^2(t') \right)\,dt'.
\end{equation}

In the original derivation \cite{Kitzler_PRL_2002},
$D(\mathbf{p})=\melement{\mathbf{p}}{z}{\Psi_0}$ was evaluated with
$\bra{\mathbf{p}}$ as a plane-wave state, which is equivalent to the so-called
strong-field approximation. This approximation was improved
\cite{Yudin_JPB_2008} by taking $\bra{\mathbf{p}}$ as a continuum eigenstate
of $\hat{H}_0$, which is known as the Coulomb-Volkov
approximation. We use a variant of this approximation \cite{Kornev_JPB_2002}.

Starting from the work \cite{Mairesse_PRA_2005,Quere_JMO_2005} of Y.\ Mairesse
and F.\ Qu\'{e}r\'{e}, the analysis of attosecond spectrograms is based on the
striking similarity between Eq.~\eqref{eq:spectrogram1} and the definition of
a spectrogram in the context of frequency-resolved optical gating (FROG):
\begin{equation}
  \label{eq:FROG_spectrogram}
  S_\mathrm{FROG}(\omega,\tau) = \left|\frac{1}{2} \int_{-\infty}^\infty dt\,
    P(t+\tau) G(t) e^{\iu \omega t} \right|^2,
\end{equation}
where $P(t)$ and $G(t)$ are interpreted as a \emph{pulse} \footnote{In the
  literature on frequency-resolved optical gating, $P(t)$ is usually referred
  to as ``probe''.} and a \emph{gate}, respectively. Powerful algorithms were
developed for retrieving both the pulse and the gate from a FROG spectrogram
\cite{Trebino_RSI_1997,Kane_JQE_1999}. The application of these algorithms to
the analysis of streaking spectrograms is hindered by the fact that
$G_0(\mathbf{p},t)$, unlike $G(t)$, depends on $\mathbf{p}$. Fortunately, this
dependence is often weak, especially if the matrix element $D(\mathbf{p})$ is
almost constant within the bandwidth of the XUV pulse. By replacing
$G_0(\mathbf{p},t)$ with
\begin{equation}
  \label{eq:gate1}
  G_1(t) = D\bigl(\mathbf{p}_0+\mathbf{A}_\mathrm{L}(t)\bigr)
  e^{\iu \Phi(\mathbf{p}_0,t)},
\end{equation}
which is known as the central momentum approximation, the streaking
spectrogram is considered as a FROG spectrogram with
$\omega=p^2/2-p_0^2/2$. Within this framework, the pulse retrieved by a FROG
algorithm was expected \cite{Sansone_Science_2006,Gagnon_APB_2008} to be
\begin{equation}
  \label{eq:pulse1}
  P_1(t) = \mathcal{E}_\mathrm{XUV}(t).
\end{equation}
In other words, FROG was thought to retrieve the complex envelope of the XUV
pulse. We find that this is \emph{generally not correct}. Given a spectrogram
defined by Eq.~\eqref{eq:spectrogram1}, FROG will rather retrieve a pulse
\begin{equation}
  \label{eq:pulse2}
  P_2(t) = \chi(t)
\end{equation}
and a gate given by Eq.\ \eqref{eq:gate2} below, which is approximately
represented by
\begin{equation}
  \label{eq:gate2_approximate}
  G_2(t) \approx e^{\iu \Phi(\mathbf{p}_0,t)}.
\end{equation}

Before we provide the mathematical background for this fact, let us illustrate
it by an example where the central momentum approximation \eqref{eq:gate1}
spectacularly breaks down. We take an artificial matrix element that roughly
models the $3s$ Cooper minimum in argon \cite{Cooper_PR_1962}:
\begin{equation}
  \label{eq:D_test}
  D(p)= p^2/2-p_0^2/2  
\end{equation}
with $p_0=1.77\ \mbox{at.\,u.}$ (choosing the central energy to
coincide with the Cooper minimum at $42.5\ \mbox{eV}$). Although $D(p)$ is a
real quantity in this example, it can be considered having a piecewise
constant phase with a discontinuity of $\pi$ at $p=p_0$. The XUV and laser
pulses are assumed to be polarized along the direction of photoelectron
detection.  Both pulses are bandwidth-limited with Gaussian envelopes:
$\mathcal{E}_\mathrm{XUV}(t) \propto \exp[-0.036 t^2]$ (shown in Fig.\
\ref{fig:pulses}) and $A_\mathrm{L}(t)=0.1 \exp[- 3.24\times 10^{-5} t^2]
\cos(0.0608 t)$. The full width at half maximum (FWHM) of the XUV intensity is
equal to 150 attoseconds. The FWHM of the laser pulse is 5~fs, its
central wavelength is 750~nm.

\begin{figure*}[htbp]
  \centering
  \raisebox{1.5in}{a)}
  \includegraphics{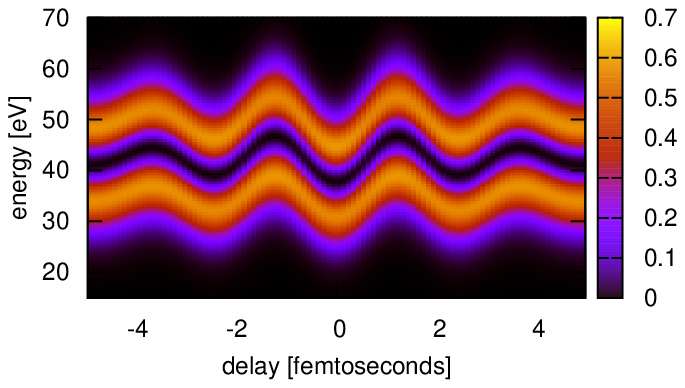}
  \raisebox{1.5in}{b)}
  \includegraphics{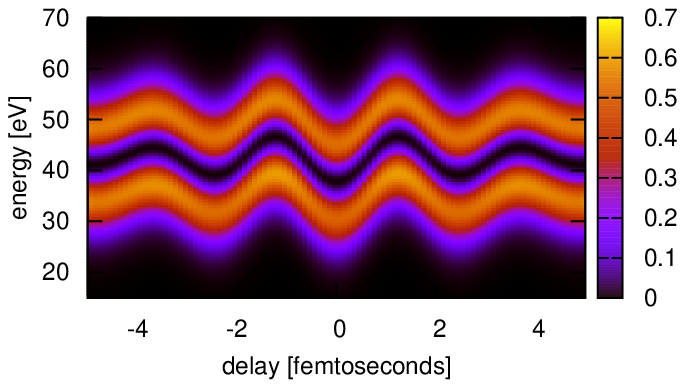}\\[-7mm]
  \raisebox{1.5in}{c)}
  \includegraphics{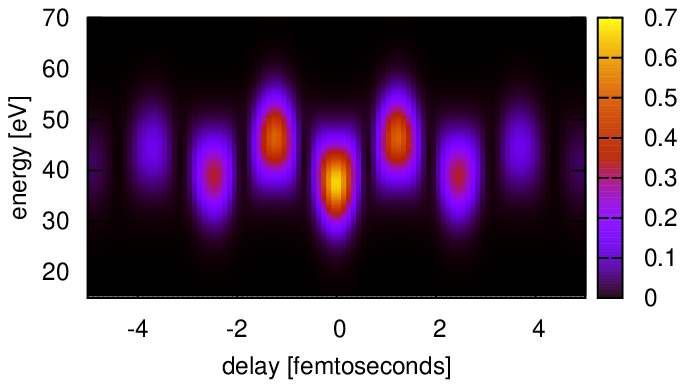}
  \raisebox{1.5in}{d)}
  \includegraphics{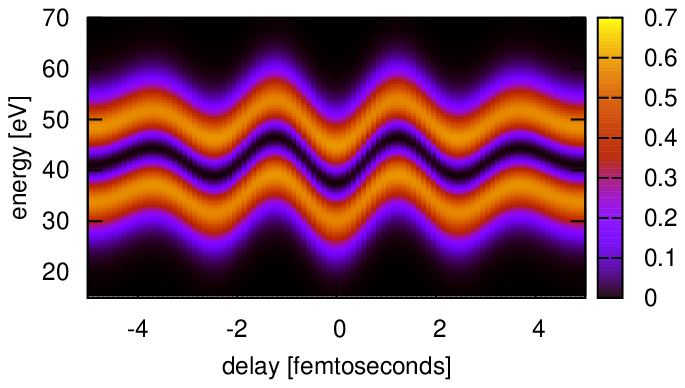}
  \caption{a) A streaking spectrogram calculated with the aid of
    Eq.~\eqref{eq:spectrogram1} using a model transition matrix element
    \eqref{eq:D_test}. b) The spectrogram reconstructed using a FROG
    algorithm. c) The spectrogram calculated using the FROG ansatz
    \eqref{eq:FROG_spectrogram} with the pulse
    $P_1(t)=\mathcal{E}_\mathrm{XUV}(t)$ and the gate $G_1(t)$. d) The FROG
    spectrogram evaluated with the pulse being the time-domain wave packet
    $\chi(t)$ and the gate $G_2(t)=\exp\{\iu \Phi(\mathbf{p}_0,t)\}$.}
  \label{fig:spectrograms}
\end{figure*}
Fig.\ \ref{fig:spectrograms}(a) shows a spectrogram calculated using
Eq.~\eqref{eq:spectrogram1}. The FROG reconstruction \cite{Gagnon_APB_2008}
yields a very similar spectrogram, Fig.\ \ref{fig:spectrograms}(b). On the other
hand, the conventional central momentum approximation, defined by
Eqs.~\eqref{eq:FROG_spectrogram}, \eqref{eq:gate1}, and \eqref{eq:pulse1},
yields a spectrogram, shown in Fig.\ \ref{fig:spectrograms}(c), which has very
little resemblance to the original spectrogram. The approximation breaks down
in this example because $D(p_0)=0$. However, the pulse and gate pair given by
Eqs.~\eqref{eq:pulse2} and \eqref{eq:gate2_approximate} yield a spectrogram
that is very close to the original one: Fig.\ \ref{fig:spectrograms}(d).

\begin{figure}[htbp]
  \centering
  \psfrag{Figure3xlabel}{$p$ \footnotesize{[atomic units]}}
  \psfrag{Figure3ylabel}{\footnotesize{radians}}
  \psfrag{Figure3label1}[][c]{$\mbox{arg}[\tilde{\chi}]$}
  \psfrag{Figure3label2}[][c]{$\mbox{arg}[\tilde{P}]$}
  \includegraphics{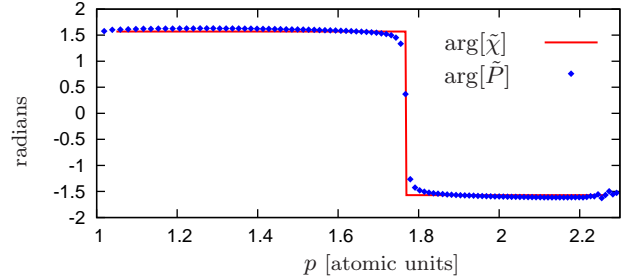}
  \caption{The phase of the momentum-space wave packet
    $\mbox{arg}[\tilde{\chi}(p)]$ (solid line) in comparison with the
    phase of the retrieved wave packet $\mbox{arg}\left[\tilde{P}\left(
        \frac{p^2}{2}-\frac{p_0^2}{2}\right)\right]$ (diamonds).}
  \label{fig:phases}
\end{figure}

In Fig.\ \ref{fig:pulses}, we compare the XUV envelope
$\mathcal{E}_\mathrm{XUV}(t)$, the time-domain wave packet $\chi(t)$, and the
pulse $P(t)$ retrieved by FROG from the spectrogram
shown in Fig.~\ref{fig:spectrograms}(a).
The retrieved pulse nearly coincides with the wave packet $\chi(t)$.
\begin{figure}[htbp]
  \centering
  \psfrag{Figure2xlabel}{$t$ \footnotesize{[atomic units]}}
  \psfrag{Figure2ylabel}{\footnotesize{arb.\ units}}
  \psfrag{Figure2label1}[][c]{$\mathcal{E}_\mathrm{XUV}(t)$}
  \psfrag{Figure2label2}[][c]{$\im[\chi(t)]$}
  \psfrag{Figure2label3}[][c]{$\im[P(t)]$}
  \includegraphics{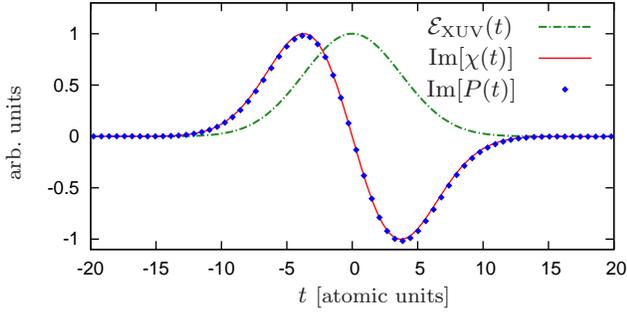}
  \caption{The envelope of the XUV pulse $\mathcal{E}_\mathrm{XUV}(t)$
    (dash-dotted line), the time-domain wave packet $\chi(t)$ defined by
    Eq.~\eqref{eq:time_domain_WP} (solid line), and the pulse $P(t)$ retrieved
    by FROG from the spectrogram shown in Fig.~\ref{fig:spectrograms}(a)
    (diamonds). In this example, the real part of the wave packet is equal to
    zero, so we compare the imaginary parts of $\chi(t)$ and $P(t)$.
  }
  \label{fig:pulses}
\end{figure}
In the momentum space, we see that the wave packet and the retrieved pulse
only differ in a narrow spectral region near the phase discontinuity, as shown in
Fig.~\ref{fig:phases}.

To gain insight as to why the time-domain wave packet
$\eqref{eq:time_domain_WP}$ plays the role of the pulse, we use
Eqs.~\eqref{eq:D} and \eqref{eq:XUV_spectrum} to express
$\mathcal{E}_\mathrm{XUV}(t+\tau)$ in \eqref{eq:spectrogram1} via
$\tilde{\chi}(p)$ and then expand $[D(p)]^{-1}$ in a Taylor series, which
eventually allows us to rewrite the master equation \eqref{eq:spectrogram1}
in a form where $\chi(t)$ plays the role of the pulse. In the case when the
XUV and streaking pulses are polarized in the direction where photoelectrons
are detected, $S(p,\tau)$ can be rewritten, without any approximations, as
\begin{equation}
  \label{eq:spectrogram2}
  S(p,\tau) = \left| \frac{1}{2}
    \int_{-\infty}^\infty dt\, \chi(t+\tau)
    \sum_{n=0}^\infty a_n \pdiffn{g(p,t)}{t}{n} \right|^2,
\end{equation}
where
$  a_n = \frac{(-i)^n}{n!}
  \evalat{\pdiffn{}{\omega}{n}
    \frac{1}{D\left(\sqrt{2\omega+p_0^2}\right)}}{\omega=0}$
and
\begin{equation}
  \label{eq:g}
  g(p,t) = D\bigl(p+A_\mathrm{L}(t)\bigr)
  e^{\iu \Phi(p,t) + \iu \left(\frac{p^2}{2}-\frac{p_0^2}{2}\right) t}.
\end{equation}

Consequently, the FROG gate associated with the time-domain wave packet
is given by
\begin{equation}
  \label{eq:gate2}
  G_2(t) = \evalat{\left(
      \sum_{n=0}^\infty a_n \pdiffn{g(p,t)}{t}{n}\right)}{p=p_0}.
\end{equation}
Applying the central momentum approximation to Eq.~\eqref{eq:gate2}, rather
than Eq.~\eqref{eq:spectrogram1}, yields a spectrogram that is closer to the
exact one. To show this analytically, we neglect all derivatives of
$A_\mathrm{L}(t)$ in the expression for $\partial^n g(p,t) / \partial t^n$,
which is justified if the exponential function in Eq.~\eqref{eq:g} oscillates
with a period that is much shorter than the optical cycle of the streaking
field. Once this approximation is made, we recognize that the sum over $n$ in
Eq.~\eqref{eq:spectrogram2} becomes the Taylor expansion of
$[D\bigl(p+A_\mathrm{L}(t)\bigr)]^{-1}$, which simplifies
Eq.~\eqref{eq:spectrogram2} to
\begin{equation}
  \label{eq:spectrogram3}
  S(p,\tau) \approx \left| \frac{1}{2}
    \int_{-\infty}^\infty dt\, \chi(t+\tau)
    e^{\iu \Phi(p,t)}
    e^{\iu \left(\frac{p^2}{2} - \frac{p_0^2}{2} \right) t} \right|^2.
\end{equation}
Thus, $G_2(t) \approx \exp\{\iu \Phi(\mathbf{p}_0,t)\}$. Importantly, the
central momentum approximation applied this way does not affect the transition
matrix element $D(p)$, while the conventional expression for the FROG gate
\eqref{eq:gate1} approximates $D(p+A_\mathrm{L}(t))$ with
$D(p_0+A_\mathrm{L}(t))$. In the example presented above, this was a very poor
approximation.

As a remark, we were able to find cases where Eq.~\eqref{eq:spectrogram3} was
a seemingly worse approximation to the true spectrogram
\eqref{eq:spectrogram1} than the conventional central momentum
approximation \footnote{For example, if we use
$D(p) = \exp\left\{4 i (p^2 - p_0^2)^2 \right\}$
instead of Eq.~\eqref{eq:D}.}.
Nevertheless, FROG retrieved the time-domain wave
packet. This confirms that the central momentum approximation is \emph{more
  generally applicable} if the spectrogram is expressed via $\chi(t)$
according to Eq.~\eqref{eq:spectrogram2}.

In conclusion, we have established that the time-domain wave packet $\chi(t)$
defined by Eq.~\eqref{eq:time_domain_WP} determines the output of attosecond
streaking measurements processed with conventional retrieval algorithms, such
as ``frequency-resolved optical gating for complete reconstruction of
attosecond bursts'' (FROG CRAB) \cite{Mairesse_PRA_2005}. The analysis
presented in this paper was focused on direct single-photon ionization
with a complex transition matrix element, using a Cooper minimum as an
example. However, it is known that Eq.~\eqref{eq:spectrogram1}
formally describes more sophisticated streaking measurements, such as those of
the Fano resonances and autoionization decay
\cite{Wickenhauser_PRL_2005,Zhao_PRA_2005}. Therefore, our analysis is
directly applicable to these situations as well, Eq.~\eqref{eq:time_domain_WP}
being a fundamental relation between the momentum-space wave function of a
photoelectron and the pulse retrieved from a streaking spectrogram.

Having established this relation, we claim that the quantum phase associated
with bound-free transitions can be measured (up to a constant phase) by means
of attosecond streaking. According to Eq.~\eqref{eq:D}, this is possible if
the spectral phase of an attosecond pulse is known. An XUV pulse can be fully
characterized in a streaking measurement performed on an atom where the
accurate matrix elements are known from a reliable model or their phases are
known to be negligible. For example, helium can be used for such a
calibration. Then another streaking measurement on the system under scrutiny
will provide all the necessary information to retrieve the unknown phases of
quantum transitions and thus completely characterize photoionization dynamics.
Such a measurement would be a test and, possibly, a challenge for
many-electron theories of photoionization, especially if the ionization
involves a resonance state or a Cooper minimum.

\begin{acknowledgments}
  Supported by the DFG Cluster of Excellence: Munich-Centre for Advanced
  Photonics.
\end{acknowledgments}


\begin{thebibliography}{10}%
\makeatletter
\providecommand \@ifxundefined [1]{%
 \ifx #1\undefined \expandafter \@firstoftwo
 \else \expandafter \@secondoftwo
\fi
}%
\providecommand \@ifnum [1]{%
 \ifnum #1\expandafter \@firstoftwo
 \else \expandafter \@secondoftwo
\fi
}%
\providecommand \enquote [1]{``#1''}%
\providecommand \bibnamefont  [1]{#1}%
\providecommand \bibfnamefont [1]{#1}%
\providecommand \citenamefont [1]{#1}%
\providecommand\href[0]{\@sanitize\@href}%
\providecommand\@href[1]{\endgroup\@@startlink{#1}\endgroup\@@href}%
\providecommand\@@href[1]{#1\@@endlink}%
\providecommand \@sanitize [0]{\begingroup\catcode`\&12\catcode`\#12\relax}%
\@ifxundefined \pdfoutput {\@firstoftwo}{%
 \@ifnum{\z@=\pdfoutput}{\@firstoftwo}{\@secondoftwo}%
}{%
 \providecommand\@@startlink[1]{\leavevmode}%
 \providecommand\@@endlink[0]{}%
}{%
 \providecommand\@@startlink[1]{%
  \leavevmode
  \pdfstartlink
   attr{/Border[0 0 1 ]/H/I/C[0 1 1]}%
   user{/Subtype/Link/A<</Type/Action/S/URI/URI(#1)>>}%
  \relax
 }%
 \providecommand\@@endlink[0]{\pdfendlink}%
}%
\providecommand \url  [0]{\begingroup\@sanitize \@url }%
\providecommand \@url [1]{\endgroup\@href {#1}{\urlprefix}}%
\providecommand \urlprefix [0]{URL }%
\providecommand \Eprint[0]{\href }%
\@ifxundefined \urlstyle {%
  \providecommand \doi [1]{doi:\discretionary{}{}{}#1}%
}{%
  \providecommand \doi [0]{doi:\discretionary{}{}{}\begingroup
  \urlstyle{rm}\Url }%
}%
\providecommand \doibase [0]{http://dx.doi.org/}%
\providecommand \Doi[1]{\href{\doibase#1}}%
\providecommand \bibAnnote [3]{%
  \BibitemShut{#1}%
  \begin{quotation}\noindent
    \textsc{Key:}\ #2\\\textsc{Annotation:}\ #3%
  \end{quotation}%
}%
\providecommand \bibAnnoteFile [2]{%
  \IfFileExists{#2}{\bibAnnote {#1} {#2} {\input{#2}}}{}%
}%
\providecommand \typeout [0]{\immediate \write \m@ne }%
\providecommand \selectlanguage [0]{\@gobble}%
\providecommand \bibinfo [0]{\@secondoftwo}%
\providecommand \bibfield [0]{\@secondoftwo}%
\providecommand \translation [1]{[#1]}%
\providecommand \BibitemOpen[0]{}%
\providecommand \bibitemStop [0]{}%
\providecommand \bibitemNoStop [0]{.\EOS\space}%
\providecommand \EOS [0]{\spacefactor3000\relax}%
\providecommand \BibitemShut [1]{\csname bibitem#1\endcsname}%
\bibitem{Drescher_Nature_2002}%
  \BibitemOpen
  \bibfield{author}{%
  \bibinfo {author} {\bibfnamefont{M.}~\bibnamefont{Drescher}}, \bibinfo
  {author} {\bibfnamefont{M.}~\bibnamefont{Hentschel}}, \bibinfo {author}
  {\bibfnamefont{R.}~\bibnamefont{Kienberger}}, \bibinfo {author}
  {\bibfnamefont{M.}~\bibnamefont{Uiberacker}}, \bibinfo {author}
  {\bibfnamefont{V.}~\bibnamefont{Yakovlev}}, \bibinfo {author}
  {\bibfnamefont{A.}~\bibnamefont{Scrinzi}}, \bibinfo {author}
  {\bibfnamefont{T.}~\bibnamefont{Westerwalbesloh}}, \bibinfo {author}
  {\bibfnamefont{U.}~\bibnamefont{Kleineberg}}, \bibinfo {author}
  {\bibfnamefont{U.}~\bibnamefont{Heinzmann}},\ and\ \bibinfo {author}
  {\bibfnamefont{F.}~\bibnamefont{Krausz}},\ }%
  \bibfield{journal}{%
  \Doi{10.1038/nature01143}{\bibinfo {journal} {Nature}}\ }%
  \textbf{\bibinfo {volume} {419}},\ \bibinfo {pages} {803} (\bibinfo {month}
  {Oct 24}\ \bibinfo {year} {2002})%
  \bibAnnoteFile{NoStop}{Drescher_Nature_2002}%
\bibitem{Swoboda_PRL_2010}%
  \BibitemOpen
  \bibfield{author}{%
  \bibinfo {author} {\bibfnamefont{M.}~\bibnamefont{Swoboda}}, \bibinfo
  {author} {\bibfnamefont{T.}~\bibnamefont{Fordell}}, \bibinfo {author}
  {\bibfnamefont{K.}~\bibnamefont{Kl\"under}}, \bibinfo {author}
  {\bibfnamefont{J.~M.}\ \bibnamefont{Dahlstr\"om}}, \bibinfo {author}
  {\bibfnamefont{M.}~\bibnamefont{Miranda}}, \bibinfo {author}
  {\bibfnamefont{C.}~\bibnamefont{Buth}}, \bibinfo {author}
  {\bibfnamefont{K.~J.}\ \bibnamefont{Schafer}}, \bibinfo {author}
  {\bibfnamefont{J.}~\bibnamefont{Mauritsson}}, \bibinfo {author}
  {\bibfnamefont{A.}~\bibnamefont{L'Huillier}},\ and\ \bibinfo {author}
  {\bibfnamefont{M.}~\bibnamefont{Gisselbrecht}},\ }%
  \bibfield{journal}{%
  \Doi{10.1103/PhysRevLett.104.103003}{\bibinfo {journal} {Phys.\ Rev.\
  Lett.}}\ }%
  \textbf{\bibinfo {volume} {104}},\ \bibinfo {pages} {103003} (\bibinfo
  {month} {Mar}\ \bibinfo {year} {2010})%
  \bibAnnoteFile{NoStop}{Swoboda_PRL_2010}%
\bibitem{Schultze_Science_2010}%
  \BibitemOpen
  \bibfield{author}{%
  \bibinfo {author} {\bibfnamefont{M.}~\bibnamefont{Schultze}}, \bibinfo
  {author} {\bibfnamefont{M.}~\bibnamefont{Fie{\ss}}}, \bibinfo {author}
  {\bibfnamefont{N.}~\bibnamefont{Karpowicz}}, \bibinfo {author}
  {\bibfnamefont{J.}~\bibnamefont{Gagnon}}, \bibinfo {author}
  {\bibfnamefont{M.}~\bibnamefont{Korbman}}, \bibinfo {author}
  {\bibfnamefont{M.}~\bibnamefont{Hofstetter}}, \bibinfo {author}
  {\bibfnamefont{S.}~\bibnamefont{Neppl}}, \bibinfo {author}
  {\bibfnamefont{A.~L.}\ \bibnamefont{Cavalieri}}, \bibinfo {author}
  {\bibfnamefont{Y.}~\bibnamefont{Komninos}}, \bibinfo {author}
  {\bibfnamefont{T.}~\bibnamefont{Mercouris}}, \bibinfo {author}
  {\bibfnamefont{C.~A.}\ \bibnamefont{Nicolaides}}, \bibinfo {author}
  {\bibfnamefont{R.}~\bibnamefont{Pazourek}}, \bibinfo {author}
  {\bibfnamefont{S.}~\bibnamefont{Nagele}}, \bibinfo {author}
  {\bibfnamefont{J.}~\bibnamefont{Feist}}, \bibinfo {author}
  {\bibfnamefont{J.}~\bibnamefont{Burgd\"{o}rfer}}, \bibinfo {author}
  {\bibfnamefont{A.~M.}\ \bibnamefont{Azzeer}}, \bibinfo {author}
  {\bibfnamefont{R.}~\bibnamefont{Ernstorfer}}, \bibinfo {author}
  {\bibfnamefont{R.}~\bibnamefont{Kienberger}}, \bibinfo {author}
  {\bibfnamefont{U.}~\bibnamefont{Kleineberg}}, \bibinfo {author}
  {\bibfnamefont{E.}~\bibnamefont{Goulielmakis}}, \bibinfo {author}
  {\bibfnamefont{F.}~\bibnamefont{Krausz}},\ and\ \bibinfo {author}
  {\bibfnamefont{V.~S.}\ \bibnamefont{Yakovlev}},\ }%
  \bibfield{journal}{%
  \Doi{10.1126/science.1189401}{\bibinfo {journal} {Science}}\ }%
  \textbf{\bibinfo {volume} {328}},\ \bibinfo {pages} {1658} (\bibinfo {month}
  {Jun 25}\ \bibinfo {year} {2010})%
  \bibAnnoteFile{NoStop}{Schultze_Science_2010}%
\bibitem{Itatani_PRL_2002}%
  \BibitemOpen
  \bibfield{author}{%
  \bibinfo {author} {\bibfnamefont{J.}~\bibnamefont{Itatani}}, \bibinfo
  {author} {\bibfnamefont{F.}~\bibnamefont{Qu\'er\'e}}, \bibinfo {author}
  {\bibfnamefont{G.~L.}\ \bibnamefont{Yudin}}, \bibinfo {author}
  {\bibfnamefont{M.~Y.}\ \bibnamefont{Ivanov}}, \bibinfo {author}
  {\bibfnamefont{F.}~\bibnamefont{Krausz}},\ and\ \bibinfo {author}
  {\bibfnamefont{P.~B.}\ \bibnamefont{Corkum}},\ }%
  \bibfield{journal}{%
  \Doi{10.1103/PhysRevLett.88.173903}{\bibinfo {journal} {Phys.\ Rev.\ Lett.}}\
  }%
  \textbf{\bibinfo {volume} {88}},\ \bibinfo {pages} {173903} (\bibinfo {month}
  {Apr}\ \bibinfo {year} {2002})%
  \bibAnnoteFile{NoStop}{Itatani_PRL_2002}%
\bibitem{Kitzler_PRL_2002}%
  \BibitemOpen
  \bibfield{author}{%
  \bibinfo {author} {\bibfnamefont{M.}~\bibnamefont{Kitzler}}, \bibinfo
  {author} {\bibfnamefont{N.}~\bibnamefont{Milosevic}}, \bibinfo {author}
  {\bibfnamefont{A.}~\bibnamefont{Scrinzi}}, \bibinfo {author}
  {\bibfnamefont{F.}~\bibnamefont{Krausz}},\ and\ \bibinfo {author}
  {\bibfnamefont{T.}~\bibnamefont{Brabec}},\ }%
  \bibfield{journal}{%
  \Doi{10.1103/PhysRevLett.88.173904}{\bibinfo {journal} {Phys.\ Rev.\ Lett.}}\
  }%
  \textbf{\bibinfo {volume} {88}},\ \bibinfo {pages} {173904} (\bibinfo {month}
  {Apr}\ \bibinfo {year} {2002})%
  \bibAnnoteFile{NoStop}{Kitzler_PRL_2002}%
\bibitem{Kienberger_Nature_2004}%
  \BibitemOpen
  \bibfield{author}{%
  \bibinfo {author} {\bibfnamefont{R.}~\bibnamefont{Kienberger}}, \bibinfo
  {author} {\bibfnamefont{E.}~\bibnamefont{Goulielmakis}}, \bibinfo {author}
  {\bibfnamefont{M.}~\bibnamefont{Uiberacker}}, \bibinfo {author}
  {\bibfnamefont{A.}~\bibnamefont{Baltuska}}, \bibinfo {author}
  {\bibfnamefont{V.}~\bibnamefont{Yakovlev}}, \bibinfo {author}
  {\bibfnamefont{F.}~\bibnamefont{Bammer}}, \bibinfo {author}
  {\bibfnamefont{A.}~\bibnamefont{Scrinzi}}, \bibinfo {author}
  {\bibfnamefont{T.}~\bibnamefont{Westerwalbesloh}}, \bibinfo {author}
  {\bibfnamefont{U.}~\bibnamefont{Kleineberg}}, \bibinfo {author}
  {\bibfnamefont{U.}~\bibnamefont{Heinzmann}}, \bibinfo {author}
  {\bibfnamefont{M.}~\bibnamefont{Drescher}},\ and\ \bibinfo {author}
  {\bibfnamefont{F.}~\bibnamefont{Krausz}},\ }%
  \bibfield{journal}{%
  \Doi{10.1038/nature02277}{\bibinfo {journal} {Nature}}\ }%
  \textbf{\bibinfo {volume} {427}},\ \bibinfo {pages} {817} (\bibinfo {month}
  {Feb 26}\ \bibinfo {year} {2004})%
  \bibAnnoteFile{NoStop}{Kienberger_Nature_2004}%
\bibitem{Quere_JMO_2005}%
  \BibitemOpen
  \bibfield{author}{%
  \bibinfo {author} {\bibfnamefont{F.}~\bibnamefont{Quere}}, \bibinfo {author}
  {\bibfnamefont{Y.}~\bibnamefont{Mairesse}},\ and\ \bibinfo {author}
  {\bibfnamefont{J.}~\bibnamefont{Itatani}},\ }%
  \bibfield{journal}{%
  \Doi{10.1080/09500340412331307942}{\bibinfo {journal} {J.\ Mod.\ Opt.}}\ }%
  \textbf{\bibinfo {volume} {52}},\ \bibinfo {pages} {339} (\bibinfo {month}
  {Jan-Feb}\ \bibinfo {year} {2005})%
  \bibAnnoteFile{NoStop}{Quere_JMO_2005}%
\bibitem{Gagnon_APB_2008}%
  \BibitemOpen
  \bibfield{author}{%
  \bibinfo {author} {\bibfnamefont{J.}~\bibnamefont{Gagnon}}, \bibinfo {author}
  {\bibfnamefont{E.}~\bibnamefont{Goulielmakis}},\ and\ \bibinfo {author}
  {\bibfnamefont{V.~S.}\ \bibnamefont{Yakovlev}},\ }%
  \bibfield{journal}{%
  \Doi{10.1007/s00340-008-3063-x}{\bibinfo {journal} {Appl.\ Phys.\ B}}\ }%
  \textbf{\bibinfo {volume} {92}},\ \bibinfo {pages} {25} (\bibinfo {month}
  {Jul}\ \bibinfo {year} {2008})%
  \bibAnnoteFile{NoStop}{Gagnon_APB_2008}%
\bibitem{Gagnon_OE_2009}%
  \BibitemOpen
  \bibfield{author}{%
  \bibinfo {author} {\bibfnamefont{J.}~\bibnamefont{Gagnon}}\ and\ \bibinfo
  {author} {\bibfnamefont{V.~S.}\ \bibnamefont{Yakovlev}},\ }%
  \bibfield{journal}{%
  \bibinfo {journal} {Opt.\ Expr.}\ }%
  \textbf{\bibinfo {volume} {17}},\ \bibinfo {pages} {17678} (\bibinfo {month}
  {Sep 28}\ \bibinfo {year} {2009})%
  \bibAnnoteFile{NoStop}{Gagnon_OE_2009}%
\bibitem{Mauritsson_PRA_2005}%
  \BibitemOpen
  \bibfield{author}{%
  \bibinfo {author} {\bibfnamefont{J.}~\bibnamefont{Mauritsson}}, \bibinfo
  {author} {\bibfnamefont{M.~B.}\ \bibnamefont{Gaarde}},\ and\ \bibinfo
  {author} {\bibfnamefont{K.~J.}\ \bibnamefont{Schafer}},\ }%
  \bibfield{journal}{%
  \Doi{10.1103/PhysRevA.72.013401}{\bibinfo {journal} {Phys.\ Rev.\ A}}\ }%
  \textbf{\bibinfo {volume} {72}},\ \bibinfo {pages} {013401} (\bibinfo {month}
  {Jul}\ \bibinfo {year} {2005})%
  \bibAnnoteFile{NoStop}{Mauritsson_PRA_2005}%
\bibitem{Yudin_JPB_2008}%
  \BibitemOpen
  \bibfield{author}{%
  \bibinfo {author} {\bibfnamefont{G.~L.}\ \bibnamefont{Yudin}}, \bibinfo
  {author} {\bibfnamefont{S.}~\bibnamefont{Patchkovskii}},\ and\ \bibinfo
  {author} {\bibfnamefont{A.~D.}\ \bibnamefont{Bandrauk}},\ }%
  \bibfield{journal}{%
  \Doi{10.1088/0953-4075/41/4/045602}{\bibinfo {journal} {J.\ Phys.\ B}}\ }%
  \textbf{\bibinfo {volume} {41}},\ \bibinfo {pages} {045602} (\bibinfo {month}
  {Feb 28}\ \bibinfo {year} {2008})%
  \bibAnnoteFile{NoStop}{Yudin_JPB_2008}%
\bibitem{Kornev_JPB_2002}%
  \BibitemOpen
  \bibfield{author}{%
  \bibinfo {author} {\bibfnamefont{A.~S.}\ \bibnamefont{Kornev}}\ and\ \bibinfo
  {author} {\bibfnamefont{B.~A.}\ \bibnamefont{Zon}},\ }%
  \bibfield{journal}{%
  \Doi{10.1088/0953-4075/35/11/304}{\bibinfo {journal} {J.\ of Phys.\ B}}\ }%
  \textbf{\bibinfo {volume} {35}},\ \bibinfo {pages} {2451} (\bibinfo {year}
  {2002})%
  \bibAnnoteFile{NoStop}{Kornev_JPB_2002}%
\bibitem{Mairesse_PRA_2005}%
  \BibitemOpen
  \bibfield{author}{%
  \bibinfo {author} {\bibfnamefont{Y.}~\bibnamefont{Mairesse}}\ and\ \bibinfo
  {author} {\bibfnamefont{F.}~\bibnamefont{Qu\'er\'e}},\ }%
  \bibfield{journal}{%
  \Doi{10.1103/PhysRevA.71.011401}{\bibinfo {journal} {Phys.\ Rev.\ A}}\ }%
  \textbf{\bibinfo {volume} {71}},\ \bibinfo {pages} {011401(R)} (\bibinfo
  {month} {Jan}\ \bibinfo {year} {2005})%
  \bibAnnoteFile{NoStop}{Mairesse_PRA_2005}%
\bibitem{Note1}%
  \BibitemOpen
  \bibinfo {note} {In the literature on frequency-resolved optical gating,
  $P(t)$ is usually referred to as ``probe''.}%
  \bibAnnoteFile{Stop}{Note1}%
\bibitem{Trebino_RSI_1997}%
  \BibitemOpen
  \bibfield{author}{%
  \bibinfo {author} {\bibfnamefont{R.}~\bibnamefont{Trebino}}, \bibinfo
  {author} {\bibfnamefont{K.~W.}\ \bibnamefont{DeLong}}, \bibinfo {author}
  {\bibfnamefont{D.~N.}\ \bibnamefont{Fittinghoff}}, \bibinfo {author}
  {\bibfnamefont{J.~N.}\ \bibnamefont{Sweetser}}, \bibinfo {author}
  {\bibfnamefont{M.~A.}\ \bibnamefont{Krumbugel}}, \bibinfo {author}
  {\bibfnamefont{B.~A.}\ \bibnamefont{Richman}},\ and\ \bibinfo {author}
  {\bibfnamefont{D.~J.}\ \bibnamefont{Kane}},\ }%
  \bibfield{journal}{%
  \bibinfo {journal} {Rev.\ Scient.\ Instr.}\ }%
  \textbf{\bibinfo {volume} {68}},\ \bibinfo {pages} {3277} (\bibinfo {month}
  {Sep}\ \bibinfo {year} {1997})%
  \bibAnnoteFile{NoStop}{Trebino_RSI_1997}%
\bibitem{Kane_JQE_1999}%
  \BibitemOpen
  \bibfield{author}{%
  \bibinfo {author} {\bibfnamefont{D.~J.}\ \bibnamefont{Kane}},\ }%
  \bibfield{journal}{%
  \bibinfo {journal} {IEEE J.\ Quant.\ Electron.}\ }%
  \textbf{\bibinfo {volume} {35}},\ \bibinfo {pages} {421} (\bibinfo {month}
  {Apr}\ \bibinfo {year} {1999}),\ ISSN \bibinfo {issn} {0018-9197}%
  \bibAnnoteFile{NoStop}{Kane_JQE_1999}%
\bibitem{Sansone_Science_2006}%
  \BibitemOpen
  \bibfield{author}{%
  \bibinfo {author} {\bibfnamefont{G.}~\bibnamefont{Sansone}}, \bibinfo
  {author} {\bibfnamefont{E.}~\bibnamefont{Benedetti}}, \bibinfo {author}
  {\bibfnamefont{F.}~\bibnamefont{Calegari}}, \bibinfo {author}
  {\bibfnamefont{C.}~\bibnamefont{Vozzi}}, \bibinfo {author}
  {\bibfnamefont{L.}~\bibnamefont{Avaldi}}, \bibinfo {author}
  {\bibfnamefont{R.}~\bibnamefont{Flammini}}, \bibinfo {author}
  {\bibfnamefont{L.}~\bibnamefont{Poletto}}, \bibinfo {author}
  {\bibfnamefont{P.}~\bibnamefont{Villoresi}}, \bibinfo {author}
  {\bibfnamefont{C.}~\bibnamefont{Altucci}}, \bibinfo {author}
  {\bibfnamefont{R.}~\bibnamefont{Velotta}}, \bibinfo {author}
  {\bibfnamefont{S.}~\bibnamefont{Stagira}}, \bibinfo {author}
  {\bibfnamefont{S.}~\bibnamefont{De~Silvestri}},\ and\ \bibinfo {author}
  {\bibfnamefont{M.}~\bibnamefont{Nisoli}},\ }%
  \bibfield{journal}{%
  \Doi{10.1126/science.1132838}{\bibinfo {journal} {Science}}\ }%
  \textbf{\bibinfo {volume} {314}},\ \bibinfo {pages} {443} (\bibinfo {month}
  {Oct 20}\ \bibinfo {year} {2006})%
  \bibAnnoteFile{NoStop}{Sansone_Science_2006}%
\bibitem{Cooper_PR_1962}%
  \BibitemOpen
  \bibfield{author}{%
  \bibinfo {author} {\bibfnamefont{J.~W.}\ \bibnamefont{Cooper}},\ }%
  \bibfield{journal}{%
  \Doi{10.1103/PhysRev.128.681}{\bibinfo {journal} {Phys.\ Rev.}}\ }%
  \textbf{\bibinfo {volume} {128}},\ \bibinfo {pages} {681} (\bibinfo {month}
  {Oct}\ \bibinfo {year} {1962})%
  \bibAnnoteFile{NoStop}{Cooper_PR_1962}%
\bibitem{Note2}%
  \BibitemOpen
  \bibinfo {note} {For example, if we use $D(p) = \protect \qopname \relax
  o{exp}\left \protect \{4 i (p^2 - p_0^2)^2 \right \protect \}$ instead of
  Eq.~\protect \textup {\hbox {\mathsurround \z@ \protect \normalfont
  (\ignorespaces \ref {eq:D}\unskip \@@italiccorr )}}.}%
  \bibAnnoteFile{Stop}{Note2}%
\bibitem{Wickenhauser_PRL_2005}%
  \BibitemOpen
  \bibfield{author}{%
  \bibinfo {author} {\bibfnamefont{M.}~\bibnamefont{Wickenhauser}}, \bibinfo
  {author} {\bibfnamefont{J.}~\bibnamefont{Burgd\"orfer}}, \bibinfo {author}
  {\bibfnamefont{F.}~\bibnamefont{Krausz}},\ and\ \bibinfo {author}
  {\bibfnamefont{M.}~\bibnamefont{Drescher}},\ }%
  \bibfield{journal}{%
  \Doi{10.1103/PhysRevLett.94.023002}{\bibinfo {journal} {Phys.\ Rev.\ Lett.}}\
  }%
  \textbf{\bibinfo {volume} {94}},\ \bibinfo {pages} {023002} (\bibinfo {month}
  {Jan}\ \bibinfo {year} {2005})%
  \bibAnnoteFile{NoStop}{Wickenhauser_PRL_2005}%
\bibitem{Zhao_PRA_2005}%
  \BibitemOpen
  \bibfield{author}{%
  \bibinfo {author} {\bibfnamefont{Z.~X.}\ \bibnamefont{Zhao}}\ and\ \bibinfo
  {author} {\bibfnamefont{C.~D.}\ \bibnamefont{Lin}},\ }%
  \bibfield{journal}{%
  \Doi{10.1103/PhysRevA.71.060702}{\bibinfo {journal} {Phys.\ Rev.\ A}}\ }%
  \textbf{\bibinfo {volume} {71}},\ \bibinfo {pages} {060702(R)} (\bibinfo
  {month} {Jun}\ \bibinfo {year} {2005})%
  \bibAnnoteFile{NoStop}{Zhao_PRA_2005}%
\end{thebibliography}
\end{document}